\documentclass[11pt,twoside]{article}
\usepackage{asp2010,natbib}

\resetcounters
\begin{document}

\title{PACS spectroscopy of OH/IR stars}
\author{R. Lombaert$^1$, B. L. de Vries$^1$, L. Decin$^{1,2}$, J.A.D.L. Blommaert$^1$, P. Royer$^1$, E. De Beck$^1$, A. de Koter$^{2,3}$, L. B. F. M. Waters$^{2,4}$
\affil{$^1$Instituut voor Sterrenkunde, K.U.Leuven, Celestijnenlaan 200D, B-3001 Leuven, Belgium}
\affil{$^2$Astronomical Institute ``Anton Pannekoek'', University of Amsterdam, Science Park XH, Amsterdam, The Netherlands}
\affil{$^3$Astronomical Institute Utrecht, University Utrecht, P.O. Box 80000, 3508 TA Utrecht, The Netherlands}
\affil{$^4$Netherlands Institute for Space Research, Sorbonnelaan 2, 3584 CA Utrecht, The Netherlands}}

\begin{abstract}
Observations of high-excitation molecular emission lines can greatly increase our understanding of AGB winds, as they trace the innermost regions of the circumstellar envelope. The PACS spectrometer on-board the Herschel Space Telescope\footnote{Herschel is an ESA space observatory with science instruments provided by European-led Principal Investigator consortia and with important participation from NASA.}, provides for the first time the spectral resolution and sensitivity necessary to trace these lines. We report on the first modelling efforts of a PACS spectral scan for the OH/IR star V669 Cas. Central to our methodology is the consistent treatment of both dust and gas by using a line radiative transfer and a continuum radiative transfer code conjointly. Water emission lines are found to be extremely sensitive to the dust-to-gas ratio, emphasizing the need of consistent modelling for dust and gas.
\end{abstract}

\section{PACS and its view}
The PACS spectrometer (\citeauthor{pog2010}~\citeyear{pog2010}) on-board the Herschel Space Telescope (\citeauthor{pil2010}~\citeyear{pil2010}), covers the wavelength range between $50\ \mu$m and $200\ \mu$m at a resolution of typically 1000. In comparison to the ISO LWS instrument, PACS offers a higher resolution and a better sensitivity. With PACS, we can observe a wide range of molecular emission lines in the innermost regions of circumstellar envelopes (CSE) of AGB stars, including CO and H$_2$O, which are major coolants and thus important for determining the thermodynamical structure of these environments.

\section{Methodology and preliminary results}
Kinematical, thermodynamical and chemical information about the circumstellar shell is provided by molecular emission lines and dust features. This information is derived through the use of two radiative transfer codes. The non-LTE line radiative transfer code, \emph{GASTRoNOoM} (\citeauthor{dec2006}~\citeyear{dec2006}), calculates the velocity, temperature and density profiles of the gas envelope, the level populations of the molecules accounted for and the emergent line profiles for the different transitions of each molecule. The continuum radiative transfer code, \emph{MCMax} (\citeauthor{min2009a}~\citeyear{min2009a}), calculates the temperature structure of the dust envelope and the final SED. In order to get a full understanding of the entire envelope around an AGB source, both modelling approaches need be used while maintaining consistency between dust and gas (see Lombaert et al., in prep).

The need for a consistent treatment of both the gas and dust components is important because of the high sensitivity of the water emission models to the dust-to-gas ratio. Modelling has shown that CO emission lines do not share this sensitivity. This behaviour is shown in Figure~1, which gives an excerpt of the PACS data of V669 Cas (full black), overlayed with two models differing only in the dust-to-gas ratio. This indicates that CO can be safely used to determine the gas temperature profile and the mass loss rate and that the water abundance profile can then be derived if the dust-to-gas ratio is constrained. Therefore, we suggest to determine the dust-to-gas ratio empirically from both gas emission (i.e. CO lines) and SED (i.e. dust continuum) modelling, with a consistent iterative treatment of both gas and dust. Consequently, one can improve the constraints on the water abundance profile.

\begin{figure}\centering
\includegraphics[height=5.7cm]{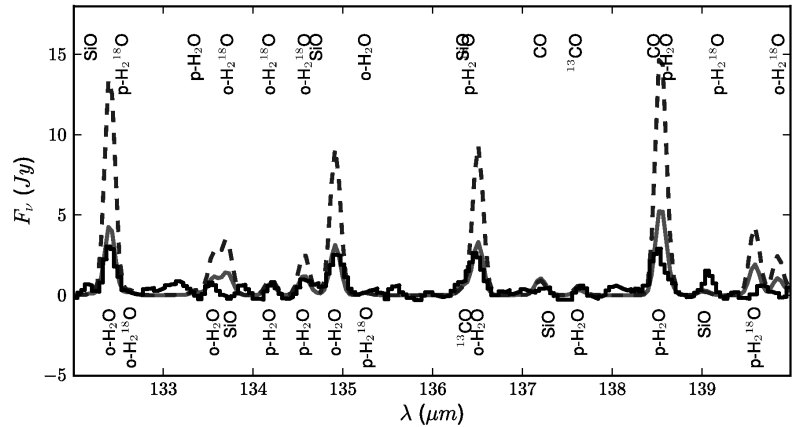}
\caption{An excerpt of the PACS spectrum of V669 Cas (full black). Two models are overplotted, as well as the molecular transitions for CO, $^{13}$CO, o-H$_2$O, p-H$_2$O, o-H$_2^{18}$O, p-H$_2^{18}$O and SiO at their expected frequencies. Model 1 (full gray) has a dust-to-gas ratio $\psi = 0.001$, whereas Model 2 (dashed) has $\psi = 0.005$.}
\end{figure}
\acknowledgements
RL acknowledges support from the KULeuven under grant number GOA-B6995, BdV and LD from the Fund for Scientific Research of Flanders (FWO), EDB from FWO under grant number G.0470.07 and JB and PR from the Belgian Federal Science Policy Office via the PRODEX Programme of ESA.
\bibliography{lombaert}

\begin{thebibliography}{}
\expandafter\ifx\csname natexlab\endcsname\relax\def\natexlab#1{#1}\fi
\expandafter\ifx\csname url\endcsname\relax
  \def\url#1{\texttt{#1}}\fi
\expandafter\ifx\csname urlprefix\endcsname\relax\def\urlprefix{URL }\fi
\providecommand{\eprint}[2][]{\url{#2}}

\bibitem[{{Decin} et~al.(2006){Decin}, {Hony}, {de Koter} et~al.}]{dec2006}
{Decin}, L., {Hony}, S., {de Koter}, A., et~al. 2006, \aap, 456, 549

\bibitem[{{Min} et~al.(2009){Min}, {Dullemond}, {Dominik} et~al.}]{min2009a}
{Min}, M., {Dullemond}, C.~P., {Dominik}, C., et~al. 2009, \aap, 497, 155

\bibitem[{{Pilbratt} et~al.(2010){Pilbratt}, {Riedinger}, {Passvogel}
  et~al.}]{pil2010}
{Pilbratt}, G.~L., {Riedinger}, J.~R., {Passvogel}, T., et~al. 2010, \aap, 518,
  L1

\bibitem[{{Poglitsch} et~al.(2010){Poglitsch}, {Waelkens}, {Geis}
  et~al.}]{pog2010}
{Poglitsch}, A., {Waelkens}, C., {Geis}, N., et~al. 2010, \aap, 518, L2

\end{thebibliography}

\end{document}